\input{aipcheck}

\documentclass[
    ,final            
  ]
  {aipproc}

\layoutstyle{6x9}

\begin{document}

\title{Multi-wavelength and black hole mass properties of Low Luminosity Active Nuclei}

\classification{98.54.Cm}
\keywords      {}

\author{F. Panessa}{
  address={Instituto de F{\'\i}sica de Cantabria (CSIC-UC),
   Avda. de los Castros, 39005 Santander, Spain}
}

\author{X. Barcons}{
  address={Instituto de F{\'\i}sica de Cantabria (CSIC-UC),
   Avda. de los Castros, 39005 Santander, Spain}
}
\author{L. Bassani}{
  address={INAF-IASF, Via P. Gobetti 101, 40129 Bologna, Italy}
}

\author{M. Cappi}{
  address={INAF-IASF, Via P. Gobetti 101, 40129 Bologna, Italy}
}

\author{F.J. Carrera}{
  address={Instituto de F{\'\i}sica de Cantabria (CSIC-UC),
   Avda. de los Castros, 39005 Santander, Spain}
}

\author{M. Dadina}{
  address={INAF-IASF, Via P. Gobetti 101, 40129 Bologna, Italy}
}

\author{L.C. Ho}{
  address={The Observatories of the Carnegie Institution of Washington, 813 Santa Barbara St.
	     Pasadena, CA 91101}
}
\author{K. Iwasawa}{
  address={Max Planck Institut f\"{u}r Extraterrestrische Physik (MPE), Giessenbachstrasse 1,
	     D-85748 Garching, Germany}
}

\author{S. Pellegrini}{
  address={Dipartimento di Astronomia, Universita' di Bologna
via Ranzani 1, 40127 Bologna, Italy}
}

\begin{abstract}

We investigate the relation between the X-ray nuclear emission,
optical emission line, radio luminosity and black hole mass
for a sample of nearby Seyfert galaxies.
Strong linear correlations between the 2-10 keV and [OIII], 
radio luminosities have been found,
showing the same slopes found in quasars and luminous Seyfert galaxies,
thus implying independence from the level of nuclear activity
displayed by the sources. Moreover, despite the wide range
of Eddington ratios (L/L$_{Edd}$) tested here (six orders
of magnitude, from 0.1 down to 10$^{-7}$),
no correlation is found between
the X-ray, optical emission lines, radio luminosities and the black hole mass.
These results suggest that low luminosity Seyfert galaxies are a scaled down version of 
luminous AGN and probably are powered by the same physical processes.
\end{abstract}

\maketitle


\section{Introduction}

One of the distinctive characteristic of nearby nuclei
is their intrinsic faintness, i.e. L$_{Bol}$ $<$ 10$^{44}$ erg/s, 
as well as their low level of activity;
in terms of Eddington luminosity most of them
have L/L$_{Edd}$ $<$ 10$^{-2}$ compared to
L/L$_{Edd}$ $\sim$ 1 of luminous AGN.
Whether low luminosity AGN (LLAGN) are a scaled-down luminosity version
of classical AGN or objects powered by different
physical mechanism is a debated issue. 
It is not clear in fact, whether
LLAGN are powered by radiatively inefficient accretion flows,
such as Advection Dominated Accretion Flows (ADAF)
and their variants (Narayan \& Yi 1994) instead of
the standard geometrically thin optically thick accretion disk
typically proposed as the accretion mechanism acting
in the central regions of luminous AGN (Shakura \& Sunyaev 1973). 
LLAGN could also represent scaled up versions of black
hole binaries in the steady-jet, hard X-ray state,
as pointed out by the scaling relations reported
in Falcke et al. (2004) and references therein.
On one hand, ADAF models are able to predict some of 
the spectral properties observed in many LLAGN,
such as the lack of the 'big blue bump' (Ho 1999).
On the other hand, some LLAGN show properties
which are common to luminous AGN, such as
the observed correlations between
optical emission lines and ionizing continuum
(Ho \& Peng 2001) or X-ray emission 
(Terashima et al. 2000, Ho et al. 2001).

X-rays and radio emission are among the most direct
evidences of nuclear activity and are, therefore, fundamental
to study the accretion processes.
It is particularly important to have a good characterization
of the spectra and determine their nuclear luminosities.

Closely related to the theoretical and observational
issues in LLAGN is the determination of the black holes masses.
The observed radiative output (e.g., X-ray luminosities)
combined with M$_{BH}$ estimates, allows us to measure the
Eddington ratios and, therefore investigate the
fundamental scaling of black hole properties
with M$_{BH}$ and accretion rate, $\dot{m}$.

We have chosen to investigate LLAGN
and their relation with luminous AGN by studying the properties
of a well defined sample of nearby Seyfert galaxies  
In this work, we focus on the X-ray, radio and [OIII] 
emission line properties. The
estimates of the central BH masses are then used to test the activity
levels of the sources. The strength of our approach resides in the capability
to correlate very accurate nuclear multi 
frequency luminosities for a homogeneous sample of nearby 
Seyfert galaxies and test them against their nuclear activity.

\section{The sample and the data}

The Seyfert sample here presented comprises
47 out of 60 Seyfert galaxies from the Palomar 
optical spectroscopic survey of nearby galaxies 
(Ho, Filippenko, \& Sargent 1995) for which X-ray data are available. 
The sources are classified as type~2 (34 out of 60), type~1 (13 out of 60), 
and "mixed" Seyfert galaxies (8),
according to their position in the optical emission line diagnostic diagrams 
(Osterbrock 1981).
The "mixed" Seyferts are placed near the boundary between
Seyfert and LINER, HII or transition classification, resulting in
a double classification (e.g., S2/T2, L2/S2, H/S2, etc.).
See Panessa et al. (2006) for a more detailed description of the sample.
For the purpose of our study
this is one of the best samples available up to now. In fact,
it offers an accurate optical classification and 
the opportunity of detecting weak nuclei. 
Finally, the sample covers a large range of AGN
luminosities ($L_{2-10 keV}$~$\sim$~10$^{37-43}$ erg s$^{-1}$) 
making it ideal for exploring possible trends with AGN power.

An homogeneous and standard X-ray data analysis has been carried out on
our selected Seyfert sample using {\it Chandra} and {\it XMM-Newton} observations. 
{\it Chandra} and {\it XMM-Newton} observations are available for 39
objects of the sample with 22 objects having 
observations with both observatories. 
The distributions of spectral parameters, in particular for type 1 objects,
are found to be within the range of values observed in luminous AGN.
The X-ray luminosities have been obtained in a homogeneous way and 
diffuse emission and/or off-nuclear sources have been excluded in order to get uncontaminated nuclear luminosities. A detailed description of the data analysis is reported in 
Cappi et al. (2006) and Panessa et al. (2006).

\section{Luminosity-Luminosity Correlations}


\begin{figure}
  \includegraphics[height=.3\textheight]{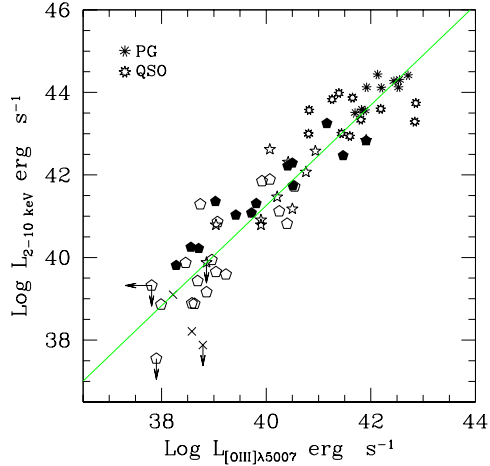}
  \caption{Log of 2-10 keV luminosity
versus log of [OIII]$_{\lambda5007}$ luminosity corrected
for the Galactic and NLR extinction. The solid line
shows the best fit linear regression line obtained by fitting
the total sample of Seyfert galaxies, the PG quasars
and bright type 1 Seyfert. Type 1 objects are plotted as filled polygons, 
type 2 as empty polygons, 'mixed Seyfert' objects
as crosses and Compton thick candidates as stars.}
\label{ox}
\end{figure}

The detection of an X-ray nucleus in almost all
our sources is a strong evidence in favour of the presence of an AGN even at very low
luminosities. The activity of the central source can also be investigated through the observed correlations between X-ray and optical emission line luminosities. In 
luminous sources strong correlations between
H$_{\alpha}$, H$_{\beta}$, [OIII]$_{\lambda5007}$ luminosities and X-ray luminosities
have been found (Mulchaey et al. 1994 and references therein). 

In Fig.~\ref{ox} we show the 2-10 keV luminosity 
(corrected taking into account the presence of 'Compton thick' sources, 
Panessa et al. 2006)
versus the [OIII]$_{\lambda5007}$ luminosity (corrected
for the Galactic and NLR extinction, Ho, Filippenko \& Sargent 1997).
Two comparison samples of bright AGN have been included
in the analysis chosen for having both X-rays and [OIII]$_{\lambda5007}$ 
fluxes available: 
1) a sample of luminous type 1 Seyfert galaxies (hereafter QSO)
from Mulchaey et al. (1994); 2)
a sample of PG quasars (hereafter PG) from Alonso-Herrero et al. (1997).
Luminosities have been adjusted to H$_{0}$= 75 km s$^{-1}$ Mpc$^{-1}$.
The two chosen samples of luminous AGN are not meant to be
complete and biases against low luminosity objects
are probably introduced. However the low luminosity ranges are covered
by our sample and they are just taken as representatives of
the class of luminous sources.

The solid line in Fig.~\ref{ox} 
is the best fit linear regression 
line obtained by fitting the Seyfert sample, the QSO and PG samples
(Log L$_{X}$ = (1.22 $\pm$ 0.06) Log L$_{[OIII]}$ + (-7.34 $\pm$ 2.53)).
The X-ray versus [OIII] correlation still holds in the flux-flux 
plot (rho=0.78, Prob $<$ 0.001). 
The [OIII]$_{\lambda5007}$ flux appears to be 
an absorption independent quantity, a good tracer of the AGN power and,
therefore, usefull to estimate the expected X-ray luminosity. 
Low Luminosity Seyfert galaxies behave like QSO and PG quasars 
suggesting a common physical nature for low and high luminosity AGN.

\begin{figure}
  \includegraphics[height=.3\textheight]{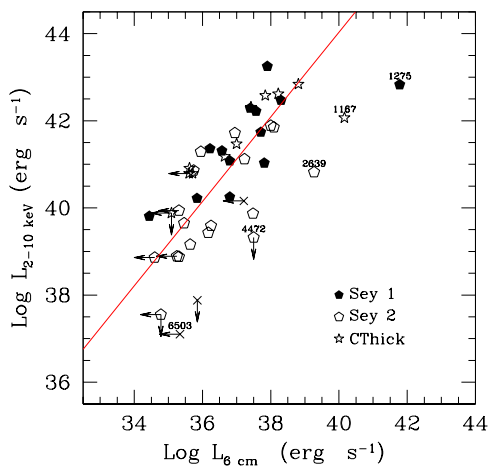}
  \caption{Intrinsic 2-10 keV luminosity
versus core radio luminosity at 6 cm. Radio data
at 6 cm are taken from Ho \& Ulvestad (2001). 
Type 1 objects are plotted as filled polygons, 
type 2 as empty polygons, 'mixed Seyfert' objects
as crosses and Compton thick candidates as stars. The NGC names
of a group of radio-loud Seyfert galaxies has been highlighted}
\label{r2}
\end{figure}

We further explore the relationship between the X-ray and 
radio luminosities, combining for the first time nuclear 
X-ray and core radio data obtained in recent surveys.
Ho \& Ulvestad (2001) have undertaken a new radio continuum 
survey of 52 Palomar Seyfert galaxies using the Very Large Array (VLA). 
The observations were made at 6 cm and at 20 cm with an 
angular resolutions of $\sim$ 1 arcsec. 
The intrinsic 2-10 keV luminosity versus the core radio luminosity at 6 cm
is plotted in Fig.~\ref{r2}.
The X-ray versus radio luminosity correlations are
highly significant (with a probability
greater than 99.9\%, using a partial Kendall $\tau$ correlation
test ). The best fit linear regression line
is overplotted in Fig.~\ref{r2} (log L$_{X}$ = (0.97$\pm$0.01) log L$_{5 GHz}$ + (5.23$\pm$0.28)).
The X-ray versus radio luminosity correlation in Fig.~\ref{r2} show
a group of outliers, i.e. sources that
show an excess in the radio emission with respect
to the average X-ray/radio ratio shown by the sample.
Interestingly, these sources have been classified in previous works
as radio-loud objects. 

The observed correlation suggests that the physics of the 
X-ray source is strongly related to the jet, where the radio emission
is thought to arise. On one hand, there is a generally accepted physical model
which predicts the X-ray emission properties in Seyfert galaxies, i.e.
in a disk-corona system the UV-soft X-ray photons from the disk
are comptonized and up-scattered into the hard X-ray band
from the hot-corona placed above the accretion disk  (Haardt \& Maraschi 1991).
On the other hand, the physics of jets in radio-quiet AGN is still unknown.
It is not clear whether the jet is characterized by relativistic or sub-relativistic speed material (Bicknell 2004),
or whether it is an aborted jet at the base of the corona (Ghisellini et al. 2004). 
The fact that the observed correlation is valid over $\sim$ 8 orders of magnitude
points to a coupling of the two components down to low luminosities.

\section{Black hole mass and accretion rates}

Black hole mass estimates are available
for 44 out of 47 objects in our sample. The M$_{BH}$ 
from the literature, have been estimated
in different ways from gas, stellar and maser kinematics
to reverberation mapping or inferred from
the mass-velocity dispersion correlations (Ferrarese 2002).

\begin{figure}
  \includegraphics[height=.3\textheight]{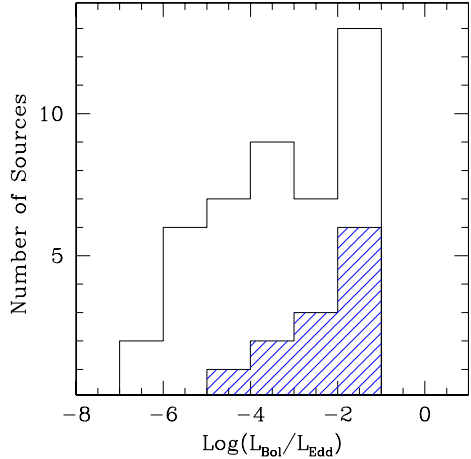}
  \caption{Distribution of the log of L$_{Bol}$/L$_{Edd}$ ratio,
assuming that L$_{Bol}$/L$_{X}$ $\sim$ 30.
In both panels the shaded areas represent the distribution of 
type 1 Seyferts only.}
\label{mbh}
\end{figure}

Black hole masses are fairly sampled 
from $\sim$ 10$^{5}$ to 10$^{8}$ M$_{\odot}$
with a peak at 10$^{7-8}$ M$_{\odot}$.
No correlation is found between the X-ray, [OIII], radio luminosity and the black hole mass. 
Some previous studies found
a correlation between the AGN luminosity and M$_{BH}$
(Kaspi et al. 2000 and references therein),
however our result is in agreement with the lack of correlation found
by Pellegrini (2005), Ho (2002) and Woo \& Urry (2002). 

The L$_{Bol}$/L$_{Edd}$ ratio
distribution is plotted in Fig.~\ref{mbh}
and it has been calculated assuming L$_{Bol}$/L$_{2-10 keV}$ $\sim$ 30.
This value is typical of luminous AGN, however it could depend
on the shape of the spectral energy distribution.
The L$_{Bol}$/L$_{Edd}$ ratio distribution for the total Seyfert sample
covers a wide range of Eddington ratios going down to 10$^{-7}$.
Woo \& Urry (2002) have shown that
bright local AGN normally show Eddington ratios
which span three orders of magnitude
down to L$_{Bol}$/L$_{Edd}$ $\sim$ 10$^{-3}$.
Indeed, most of our sources are radiating 
at very low Eddington ratios if compared with luminous AGN.
The low Eddington ratios observed in our sample are even lower
if the bolometric correction considered is that of LLAGN.
At such low Eddington ratios, radiatively inefficient accretion is
normally invoked as the putative mechanism for the production of the observed emission.
As a matter of fact, ADAF models work in a radiatively
inefficient regime at
sub-Eddington ratios (L $<$ 0.01L$_{Edd}$)
and can reproduce the lack of UV excess observed
in the SED of LLAGN.
However, also radiatively efficient standard accretion disks
are stable at low Eddington ratios down to L $\sim$ 10$^{-6}$L$_{Edd}$
(Park \& Ostriker 2001) and probably could reproduce the shape
of the LLAGN SED since the temperatures of a multi-colour disk
scale with $\dot{m}$$^{1/4}$ (Ptak et al. 2004).
Actually, the X-ray versus optical emission line and radio luminosity
correlations scale with luminosity, so that
low luminosity Seyfert galaxies appear to be a scaled-down version of
classical AGN. 


\section{Conclusions}

In the effort of further verifying the physical continuity
between our sample of Seyfert galaxies and bright AGN, the X-ray luminosities
have been correlated with the [OIII] and radio luminosities
both suspected to be good tracers of the nuclear emission. These luminosities
have also been correlated with M$_{BH}$. 
Both L$_{X}$ vs. L$_{[OIII]}$ and
L$_{X}$ vs. L$_{5 GHz}$  correlations are highly significant in our sample,
indicating that the X-ray emission and the UV ionizing radiation are linked
as well as the source of the radio emission, possibly a jet. 
Moreover, both correlations scale with luminosity over 8 orders of magnitude, suggesting that low luminosity Seyfert galaxies are powered by the same physical processes which operate in brighter AGN such as QSOs. No correlation is found between nuclear multi-band luminosities
and M$_{BH}$ in agreement with some previous studies 
(Woo \& Urry 2002, Pellegrini et al. 2005, Ho 2002).
Finally, L/L$_{Edd}$ ratios span three orders of magnitude
down to L$_{Bol}$/L$_{Edd}$ $\sim$ 10$^{-7}$, indicating
that most of our sources are accreting at very low Eddington ratios.
Overall our results suggest that Seyfert nuclei are consistent with being a scaled-down version of 
luminous AGN.




\end{document}